\documentclass[pra,onecolumn,floatfix,superscriptaddress,longbibliography]{revtex4-1}

\usepackage[T1]{fontenc}
\usepackage{natbib}
\usepackage{graphicx}
\usepackage{indentfirst}
\usepackage{listings}
\usepackage{braket}
\usepackage{amsmath}
\usepackage{amssymb}
\usepackage{amsthm}
\usepackage[font=small,labelfont=bf]{caption}
\usepackage{float}
\usepackage{subfig}
\usepackage{array, makecell}
\usepackage{bbold}
\usepackage{qcircuit}
\usepackage{placeins}
\DeclareMathOperator{\Tr}{Tr}
\usepackage[title]{appendix}
\usepackage[export]{adjustbox}
\newtheorem{definition}{Definition}
\newtheorem{theorem}{Theorem}

\begin{document}

\title{Exact analytical relation between the entropy and the dominant eigenvalue of random reduced density matrices}

\author{Ruge Lin}
\affiliation{Quantum Research Centre, Technology Innovation Institute, Abu Dhabi, UAE.}
\affiliation{Departament de F\'isica Qu\`antica i Astrof\'isica and Institut de Ci\`encies del Cosmos (ICCUB), Universitat de Barcelona, Mart\'i i Franqu\`es 1, 08028 Barcelona, Spain.}

\begin{abstract}

In this paper, we show how the entropy (including the von Neumann entropy obtained by tracing across various sizes of subsystems, the entanglement gap, as well as different degrees of R\'{e}nyi entropy) of the random reduced density matrices are related to their dominant eigenvalue. Analytical results are deduced from Random Matrix Theory (RMT) for decentralized Wishart matrices and backed up by computer simulations. The correlation between our study and entanglement generated by quantum computing is provided with various examples.

\end{abstract}

\maketitle

\section{Introduction}

The entropy of certain states is of particular importance to physicists dealing with finite-size quantum systems. More broad studies of the entropy's behavior, such as those conducted on a collection of random reduced density matrices, might be both intriguing and fruitful. 
These items may be earned in two different ways. Random density matrices may be generated using the first technique, in which the larger random state is partially traced \cite{braunstein1996geometry,sommers2004statistical}. The volume element associated with a certain distance on the set of all density matrices is used in the second technique. The first technique is the one we will discuss in this article. It is proven that the law of a density matrix (also a reduced density matrix) is the law of a Wishart matrix (defined in the preliminary) conditioned by trace $1$ \cite{nechita2007asymptotics,majumdar2010extreme}.

Therefore, we can use the random matrix theory defined on the Wishart ensemble to analyze a random reduced density matrix. The isolated dominant eigenvalue when the coefficients in the reduced density matrix do not center at $0$ can be separately analyzed using a basic theorem in linear algebra. This method mainly focuses on the expectation of the dominant eigenvalues rather than their distribution. It is sensitive to the condition that we generate random reduced density matrices (by setting the mean and the variance of the random variables). Also, it has a more straightforward analytical expression that helps us derive the relation between the entropy and the dominant eigenvalue.

\section{Preliminary}

\subsection{Marchenko Pastur distribution}

We introduce the Marchenko Pastur distribution \cite{nechita2007asymptotics,feier2012methods,livan2018introduction,bai1999kmethodologies,haagerup2003random}, which describes the asymptotic behavior of eigenvalues of a Wishart matrix. The theorem is named after mathematicians Vladimir Marchenko and Leonid Pastur, who proved this result in 1967 \cite{marvcenko1967distribution}.

\begin{definition}{Wishart matrix}

Let $X$ be an $\alpha \times \beta$ complex matrix whose elements are complex normal $\mathcal{N}_{\mathbb{C}}\left(0,1\right)$ random variables that are independently identically distributed (i.i.d). The $\alpha\times\alpha$ matrix $XX^{\dagger}$ is referred to as a Wishart randomly generated matrix with parameters $\alpha$ and $\beta$. All Wishart matrices form the Wishart ensemble \cite{wishart1928generalised}.

\end{definition}

\begin{definition}{Empirical spectral distribution, (ESD)}

Given an $\alpha\times\alpha$ symmetric or Hermitian matrix $M_{\alpha}$, consider its $\alpha$ real eigenvalues $\lambda_0, ..., \lambda_{\alpha-1}$. To study their distribution, we form the empirical spectral distribution,

\begin{equation}
\mu_{\alpha}:=\frac{1}{\alpha}\sum_{i=0}^{\alpha-1}\delta_{\lambda_i},
\end{equation}

with $\delta_{\lambda_i}\left(x\right)$ being the indicator function $\mathbb{1}_{\lambda_i\leq x}$ \cite{anderson2010introduction}.

\end{definition}

\begin{theorem}{Marchenko Pastur distribution, (MPD)}

If $X$ denotes an $\alpha \times \beta$ random matrix whose entries are i.i.d $\mathcal{N}_{\mathbb{C}}\left(0,\sigma\right)$ random variables. Let $Y_{\beta}:=\frac{1}{\beta}XX^{\dagger}$. It is a Wishart matrix. And let $\lambda_1, ..., \lambda_{\alpha}$ be eigenvalues of $Y_{\beta}$. Assume that $\alpha,\beta\rightarrow\infty$ and the ratio $\alpha/\beta\rightarrow\lambda\in\left(0,+\infty\right)$.

Then the ESD of $Y_{\beta}$, $\mu_{\alpha}$ converges weakly, in probability, to the distribution with density function $\mu$ given by

\begin{equation}
\mu =
\begin{cases}
\left(1-\frac{1}{\lambda}\right)\mathbb{1}_{0}+\nu & \text{if $\lambda > 1$} \\
\nu & \text{if $0 < \lambda \leq 1$}\\
\end{cases}\,,
\label{eq_mu_MPD}
\end{equation}

and

\begin{equation}
d\nu(x)=\frac{1}{2\pi\sigma^2} \frac{\sqrt{\left(\lambda_{+} -x\right)\left(x-\lambda_{-}\right)}}{\lambda x} \mathbb{1}_{x\in[\lambda_-,\lambda_+]}dx\,,
\label{eq_nu_MPD}
\end{equation}

with

\begin{equation}
\lambda_{\pm}=\sigma^2\left(1\pm \sqrt{\lambda}\right)^2\,.
\label{eq_lambda_MPD}
\end{equation}
\end{theorem}

The Ref. \cite{feier2012methods} contains the proof for the MPD using the moment approach. While the MPD was created and validated using the Wishart ensemble, the output is extensible.

\begin{figure}
\centering
\includegraphics[width=0.6\linewidth]{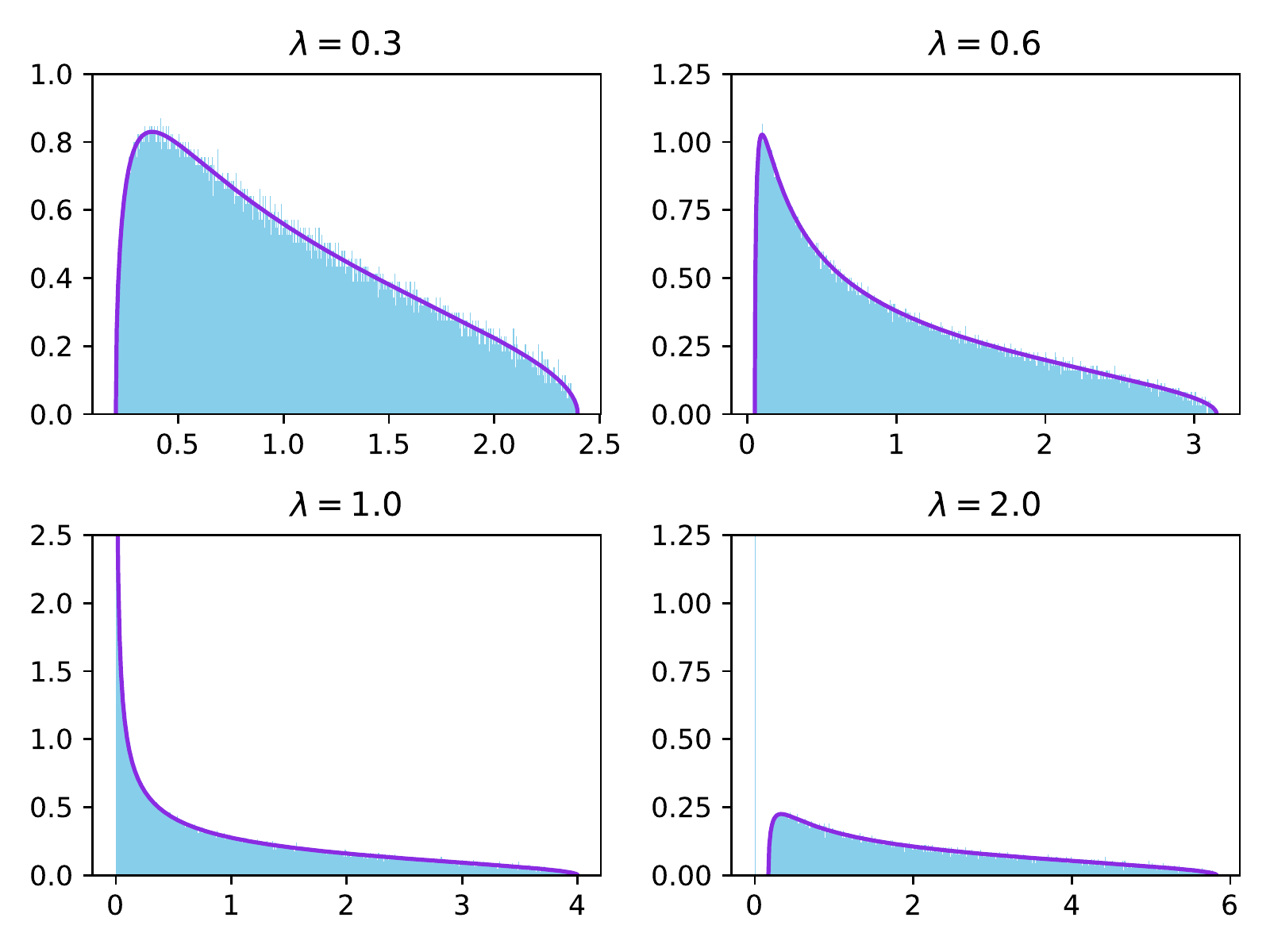}
\caption{The Marchenko–Pastur distribution (in purple) vs numerical result (in blue) for different $\lambda$ and the same $\alpha=10000$.}
\label{MPD}
\end{figure}

Let $X$ be an $\alpha\times \beta$ random matrix with variables $\mathcal{N}(0,1)$, $\lambda=\alpha/\beta$. Then the normalized histogram of the eigenvalues of $Y_{\beta}=\frac{1}{\beta}XX^{\dagger}$ is approximated by $\left(x,\frac{1}{2\pi} \frac{\sqrt{\left(\lambda_{+} -x\right)\left(x-\lambda_{-}\right)}}{\lambda x}\right)$ with $\lambda_{\pm}=\left(1\pm \sqrt{\lambda}\right)^2$ as shown in the FIG. \ref{MPD}. The code which generates every figure in this article is on GitHub \cite{Github}. 

\subsection{Dominant eigenvalue of decentralised Wishart matrices}

If $X$ denotes an $\alpha \times \beta$ complex matrix whose entries are i.i.d random variables with mean $\gamma\neq 0$ and variance $\sigma^2$, let $Y_{\beta}=\frac{1}{\beta}XX^{\dagger}$. We call $Y_\beta$ a decentralised Wishart matrix. And let $\lambda_0, ..., \lambda_{\alpha-1}$ eigenvalues of $Y_{\beta}$, with $\lambda_0 \geq ... \geq \lambda_{\alpha-1}$.

We can rewrite $X$ into

\begin{equation}
X=H+\widetilde{X},
\end{equation}

and

\begin{equation}
X^\dagger=H^\dagger+\widetilde{X}^\dagger,
\end{equation}

where $H$ is an $\alpha\times\beta$ with all elements $\gamma$, and $\widetilde{X}$ an $\alpha\times\beta$ i.i.d random matrix with mean $0$ and variance $\sigma^2$.

And we also have the relation

\begin{equation}
H=X+\left(-\widetilde{X}\right),
\end{equation}

and

\begin{equation}
H^\dagger=X^\dagger+\left(-\widetilde{X}^\dagger\right),
\end{equation}

since the elements of $\widetilde{X}$ are centred at $0$.

We define the spectral norm \cite{meyer2000matrix, belitskii2013matrix} of a matrix $P$ as

\begin{equation}
||P||:=\sqrt{\lambda_{max}\left(P^\dagger P\right)},
\end{equation}

which satisfies the inequality $||P+Q||\leq ||P||+||Q||$.

Therefore, we have

\begin{equation}
||X^\dagger||\leq ||H^\dagger|| + ||\widetilde{X}^\dagger||,
\end{equation}

which means

\begin{equation}
\sqrt{\lambda_{max}\left(XX^\dagger\right)}\leq\sqrt{\lambda_{max}\left(HH^\dagger\right)}+\sqrt{\lambda_{max}\left(\widetilde{X}\widetilde{X}^\dagger\right)}.
\end{equation}

The same way,

\begin{equation}
||H^\dagger||=||X^\dagger||+||-\widetilde{X}^\dagger||,
\end{equation}

implies

\begin{equation}
\sqrt{\lambda_{max}\left(HH^\dagger\right)}\leq\sqrt{\lambda_{max}\left(XX^\dagger\right)}+\sqrt{\lambda_{max}\left(\widetilde{X}\widetilde{X}^\dagger\right)}.
\end{equation}

So we have when $\lambda_{max}\left(HH^\dagger\right)\gg\lambda_{max}\left(\widetilde{X}\widetilde{X}^\dagger\right)=\lambda_+$, $\lambda_{max}\left(XX^\dagger\right)\sim\lambda_{max}\left(HH^\dagger\right)$.

Then $HH^\dagger$ is an $\alpha\times\alpha$ matrix with all elements equal to $\beta|\gamma|^2$, which means all its eigenvalues are $\alpha\beta|\gamma|^2$.

\begin{figure}
\centering
\includegraphics[width=0.5\linewidth]{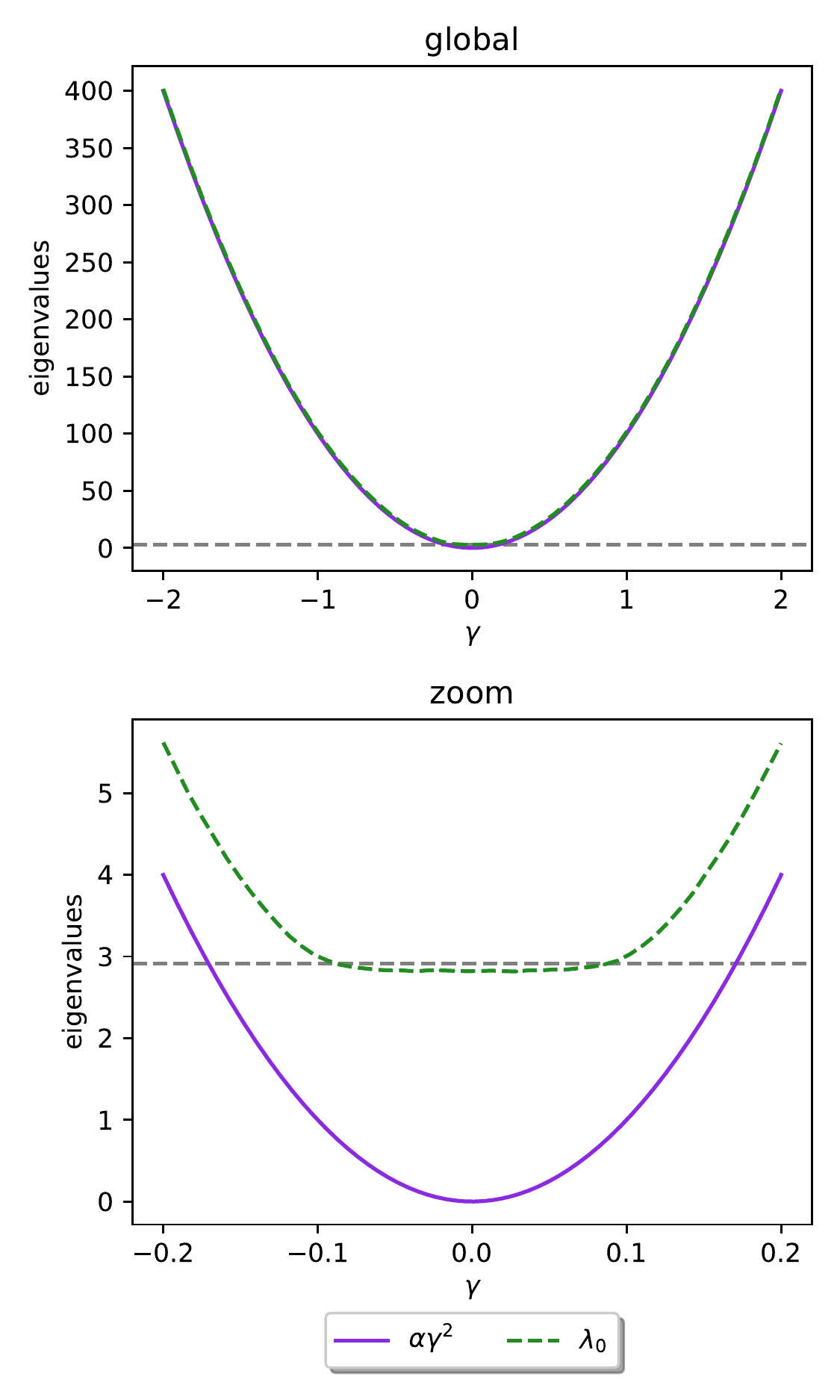}
\caption{The convergence of $\lambda_0$ towards $\alpha\gamma^2$ as $\alpha=100$, $\beta=200$ and $\gamma$ varies from $0$ to $2$. Each $\lambda_0$ is the mean of the dominant eigenvalues over $500$ random matrices $Y_\beta$ generated by the entries of $X$ following the normal distribution $\mathcal{N}\left(\gamma,1\right)$. We have $\lambda=\frac{1}{2}$, and $\lambda_+=\left(1+\frac{\sqrt{2}}{2}\right)^2$ is plotted in gray dotted line. The minor difference lies between $\lambda_0$ and $\lambda_+$ when $\gamma\sim 0$ is due to the effect of the Tracy-Widom distribution \cite{tracy1996orthogonal,tracy2002distribution,johnstone2001distribution,chiani2014distribution}. Since this phenomena is highly non analytical, also numerically negligible for our study, we will not elaborate in details.}
\label{mean_MPD}
\end{figure}

Finally we have $\lambda_0=\lambda_{max}\left(Y_\beta\right)=\lambda_{max}\left(\frac{1}{\beta}XX^\dagger\right)=\alpha|\gamma|^2$, as shown in the FIG. \ref{mean_MPD}. In the following we keep the notation $\gamma^2$ instead of $|\gamma|^2$ for simplicity.

\section{Average entropy of a subsystem}

Here we consider random density matrices obtained by partially tracing larger random pure states. There is a strong connection between these random density matrices and the Wishart ensemble of random matrix theory, that can be used to determine the formula of entropy by Don Page in 1993 \cite{nechita2007asymptotics,page1993average,bengtsson2017geometry,sen1996average, pal2020probing, zyczkowski2001induced,hayden2006aspects}.

\begin{theorem}{Average entropy of a subsystem}

If a quantum system of Hilbert space dimension $\alpha\beta$ is in a random pure state, the average von Neumann entropy $E$ of a subsystem of dimension $\alpha \leq \beta$ is $\ln \left(\alpha\right)-\frac{\alpha}{2\beta}$ for $1 \leq \alpha \leq \beta$.

\end{theorem}

We can use the MPD to reproof it.

\begin{proof}

Consider the random reduced density matrix under the form $\rho_A:=ZZ^{\dagger}/\Tr\left(ZZ^{\dagger}\right)$ \cite{nechita2007asymptotics}, where $Z$ is an $\alpha\times\beta$ random matrix for $1 \leq \alpha \leq \beta$. We have $E=\sum^{\alpha-1}_{i=0}-\lambda_{i}\ln\left(\lambda_{i}\right)$ where $\lambda_{i}$ are eigenvalues of $\rho_A$.

We can eliminate the trace and rewrite $\rho_A$ into a form that the MPD is applicable: $\rho_A=\frac{1}{\beta}XX^{\dagger}=\left(\frac{X}{\sqrt{\beta}}\right)\left(\frac{X}{\sqrt{\beta}}\right)^{\dagger}$, where $X$ is a $\alpha\times\beta$ random matrix with mean $0$ and $\sigma_0=\frac{1}{\sqrt{\alpha}}$. Therefore, $\frac{X}{\sqrt{\beta}}$ can be considered a "normalized" matrix, with coefficients $c_i$ that correspond to the quantum state $\ket{\psi}=\sum_{i=1}^{\alpha\beta}c_i\ket{i}$ with $\sum_{i=1}^{\alpha\beta}|c_i|^2=1$. We define $\lambda=\alpha/\beta$. The von Neumann entropy of a subsystem of dimension $\alpha$ can be written as

\begin{equation}
E=\alpha\int-x\ln\left(x\right)d\nu(x),
\label{eq_S_alpha_beta}
\end{equation}

with

\begin{equation}
d\nu\left(x\right)=\frac{1}{2\pi\sigma_0^2}\frac{\sqrt{\left(\lambda_{+}-x\right)\left(x-\lambda_{-}\right)}}{\lambda x}\mathbb{1}_{x\in[\lambda_-,\lambda_+]}dx,
\label{eq_dnu_ENT}
\end{equation}

and

\begin{equation}
\lambda_{\pm}=\sigma_0^2\left(1\pm \sqrt{\lambda}\right)^2.
\label{eq_lambda_ENT}
\end{equation}

Then we have

\begin{equation}
E=\frac{\alpha}{2\pi\sigma_0^2\lambda}\int^{\lambda_+}_{\lambda_-}-\ln\left(x\right)\sqrt{\left(\lambda_{+} -x\right)\left(x-\lambda_{-}\right)}dx,
\label{eq_E_alpha_beta_2}
\end{equation}

which can be simplified back into

\begin{equation}
E=\ln\left(\alpha\right)-\frac{\alpha}{2\beta},
\label{eq_S_alpha_beta_3}
\end{equation}

with the integral in the Eq. (\ref{eq_A2})

\begin{widetext}
\begin{equation}
\int^b_a \ln\left(x\right) \sqrt{\left(b-x\right)\left(x-a\right)}dx=\frac{\pi}{16}\left(a^2+6ab+b^2-4\sqrt{ab}\left(a+b\right)-4\left(a-b\right)^2\ln\left(2\right)+2\left(a-b\right)^2\ln\left(a+b+2\sqrt{ab}\right)\right).
\label{eq_A2}
\end{equation}
\end{widetext}

\end{proof}

\section{Average entropy of a subsystem (general case)}\label{general}

In this section, we consider a slightly different variant when the mean of $X$ is not $0$ but $\gamma$. We still have $\lambda=\alpha/\beta$ and the eigenvalues of $\rho_A$ are still $\lambda_0, ..., \lambda_{\alpha-1}$. This time we have $\lambda_0 \geq ... \geq \lambda_{\alpha-1}$ and $\lambda_0$ is the dominant eigenvalue. The standard deviation $\sigma_\gamma$ should be evaluated with $\gamma$ to maintain the sum of all eigenvalues as $1$. We have

\begin{equation}
\lambda_0=\alpha\gamma^2.
\end{equation}

The remaining eigenvalues $\lambda_1, ... ,\lambda_{\alpha-1}$ will follow the MPD, and their sum would be

\begin{equation}
\sum^{\alpha-1}_{i=1}\lambda_i\sim\alpha\int x d\nu(x)=\frac{\alpha}{2\pi\lambda\sigma_{\gamma}^2} \int_{\lambda_{-}}^{\lambda_{+}}\sqrt{\left(\lambda_{+}-x\right)\left(x-\lambda_{-}\right)}dx=\alpha\sigma_{\gamma}^2,
\label{eq_dnu_ENT}
\end{equation}

with the integral in the Eq. (\ref{eq_A1})

\begin{equation}
\int^b_a \sqrt{\left(b-x\right)\left(x-a\right)}dx=\frac{\pi}{8}\left(b-a\right)^2.
\label{eq_A1}
\end{equation}

To have the sum of all eigenvalues of $\rho_A$ equals to $1$, we should have the relation

\begin{equation}
\alpha\left(\gamma^2+\sigma_{\gamma}^2\right)=1.
\label{eq_sum1}
\end{equation}

This relation can also be obtained using the assumption that the quantum state is normalized, or $\rho_A$ has the trace $1$. These three conditions are equivalent, and all originated from the fact that only pure quantum states are investigated in the article. Here we highlight that this is the only element of quantum physics that appears in the main text. Further studies of quantum states that emerged from quantum computing are provided in the appendices.

We can deduce that

\begin{equation}
\sigma_{\gamma}^2=\frac{1-\lambda_0}{\alpha},
\label{eq_variance}
\end{equation}

then use this expression to replace $\sigma_0$ in the Eq. (\ref{eq_lambda_ENT}) and the Eq. (\ref{eq_E_alpha_beta_2}). We then have the entropy of a subsystem depending on $\lambda_0$

\begin{equation}
\begin{aligned}
E_{\lambda_0}=&\frac{\alpha}{2\pi\sigma_{\gamma}^2\lambda}\int^{\lambda_+}_{\lambda_-}-\ln\left(x\right)\sqrt{\left(\lambda_{+} -x\right)\left(x-\lambda_{-}\right)}dx-\lambda_0\ln\left(\lambda_0\right),
\end{aligned}
\label{eq_E_alpha_beta_lambda_2}
\end{equation}

with

\begin{equation}
\lambda_{\pm}=\sigma_{\gamma}^2\left(1\pm \sqrt{\lambda}\right)^2.
\label{eq_lambda_ENT_2}
\end{equation}

After applying the integral in the Eq. (\ref{eq_A2}) and massive reduction, we finally have

\begin{widetext}
\begin{equation}
E_{\lambda_0}=\frac{1}{4\alpha\beta}\left(
\begin{aligned}
& -4\alpha\beta\lambda_0\ln\left(\lambda_0\right)+\left(8\alpha\beta\ln\left(2\right) - \alpha^2 - \beta^2\right)\left(1-\lambda_0\right)\\
& -4\alpha\beta\left(1-\lambda_0\right)\ln\left(-2\left(\left(\alpha + \beta\right)\lambda_0 - \left(\beta-\alpha\right)\left(1-\lambda_0\right) - \alpha - \beta\right)/\alpha\beta\right)\\
& +\left(\alpha+\beta\right)\left(\beta-\alpha\right)\left(1-\lambda_0\right)
\end{aligned}
\right).
\label{eq_E_alpha_beta_lambda_3}
\end{equation}
\end{widetext}

With this expression in the Eq. (\ref{eq_E_alpha_beta_lambda_3}) we can recover the Eq. (\ref{eq_E_alpha_beta_2}) for $\lambda_0\rightarrow 0$. When $\beta\rightarrow\infty$, the relation between $\alpha$, $\lambda_0$ and the entropy becomes

\begin{equation}
\begin{aligned}
\lim_{\beta\rightarrow\infty}E_{\lambda_0}=&\left(-2\ln\left(2\right)-\ln\left(\alpha\right)-\ln\left(\lambda_0\right)\right)\lambda_0+\left(\lambda_0-1\right)\ln\left(4-4\lambda_0\right)+2\ln\left(2\right)+\ln\left(\alpha\right).
\end{aligned}
\end{equation}

\begin{figure}
\centering
\includegraphics[width=0.5\linewidth]{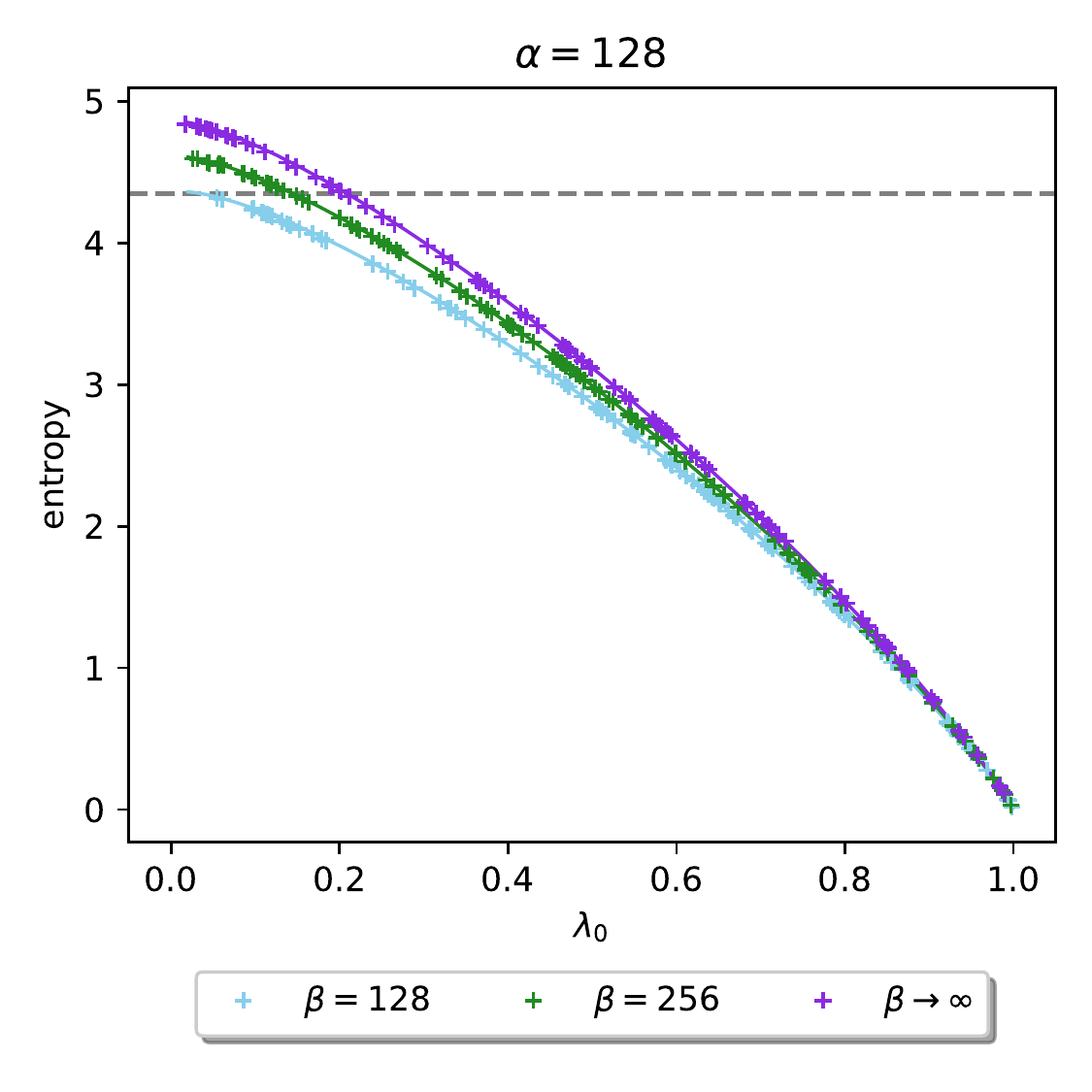}
\caption{The numerical result (in $+$) versus the analytical solution (in line). The von Neumann entropy of different $\beta>\alpha$ for the same $\alpha=128$. Here for the numerical result of $\beta\rightarrow\infty$ we use $\beta=8192$. Each of the $300$ data points in the plot is the entropy a randomly generated normalized reduced density matrix under the form $\rho_A=ZZ^{\dagger}/\Tr\left(ZZ^{\dagger}\right)$. The gray dotted line is $\ln\left(\alpha\right)-\frac{1}{2}$.}
\label{alphabeta}
\end{figure}

In the FIG. \ref{alphabeta} we compare our analytical solution with the numerical result. Though the outcome is deduced from RMT with the study of Wishart matrices, it can be extended to an actual physical system; see Appendix \ref{app_QFT} and \ref{app_grover}.

\section{Entanglement Gap}\label{Gap}

In this section, we investigate the relation between $\lambda_0$ and the entanglement gap. The reduced density matrix $\rho_A$ can be written as

\begin{equation}
\rho_A=e^{-\mathcal{H}_A},
\end{equation}

where $\mathcal{H}_A$ is called the entanglement Hamiltonian \cite{li2008entanglement,cirac2011entanglement,roy2021bulk,alba2021entanglement,calabrese2008entanglement,lauchli2010disentangling}. The entanglement spectrum levels $\xi_i=-\ln\left(\lambda_i\right)$ are the "energies" of $\mathcal{H}_A$. The entanglement gap is a natural quantity defined as

\begin{equation}
\delta\xi:=\xi_1-\xi_0.
\end{equation}

To calculate $\delta\xi$ analytically, we take $\lambda_+$ the right end of the MPD as $\lambda_1$, which is $\sigma_\gamma^2\left(1+\sqrt{\lambda}\right)^2$, and $\delta\xi$ can be written as a function of $\alpha$, $\beta$ and $\lambda_0$ with the Eq. (\ref{eq_variance})

\begin{equation}
\begin{aligned}
\delta\xi&=\ln\left(\lambda_0\right)-\ln\left(\sigma_\gamma^2\left(1+\sqrt{\lambda}\right)^2\right)=\ln\left(\lambda_0\right)-\ln\left(\frac{1-\lambda_0}{\alpha}\left(1+\sqrt{\frac{\alpha}{\beta}}\right)^2\right),
\end{aligned}
\label{eq_gap}
\end{equation}

and shown as in the FIG. \ref{Gap}. When $\beta\rightarrow\infty$, we have

\begin{equation}
\lim_{\beta\rightarrow\infty}\delta\xi=\ln\left(\lambda_0\right)-\ln\left(\frac{1-\lambda_0}{\alpha}\right).
\end{equation}

\begin{figure}
\centering
\includegraphics[width=0.5\linewidth]{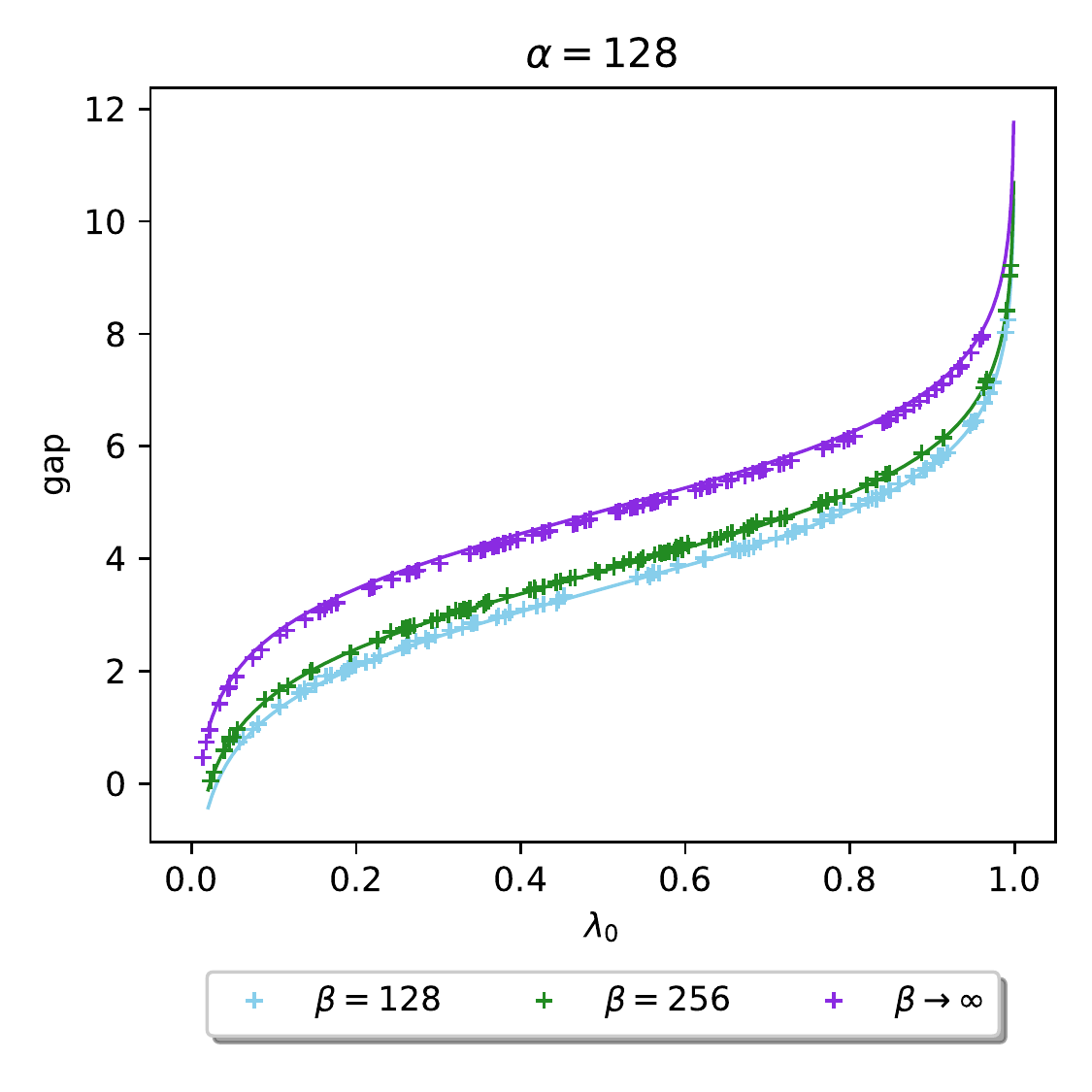}
\caption{The numerical simulation (in $+$) versus the analytical solution (in line). The entanglement gap for $300$ random reduced density matrices and the Eq. (\ref{eq_gap}). Here for the numerical result of $\beta\rightarrow\infty$ we use $\beta=2^{16}$.}
\label{Gap}
\end{figure}

From the FIG. \ref{alphabeta} and the FIG. \ref{Gap}, we can observe a monotone decreasing relation between $\lambda_0$ and $E_{\lambda_0}$, and a monotone increasing relation between $\lambda_0$ and $\delta\xi$. These two relations imply that $E_{\lambda_0}$ decreases with $\delta\xi$, as the result stated by Li and Haldane in 2008 with the ideal Moore-Read state \cite{li2008entanglement}.

\section{Different degrees of R\'{e}nyi Entropy of Bi-partition Subsystems}

In this section, we take $\beta=\alpha$ and study the relation between $\lambda_0$ and different degrees of R\'{e}nyi entropy bi-partition subsystems.

The R\'{e}nyi entropy is defined as \cite{nielsen2002quantum}

\begin{equation}
E_d:=\frac{1}{1-d}\ln\left(\sum^{\alpha-1}_{i=0}\lambda_i^d\right),
\end{equation}

where $d$ is the degree, and $\lambda_i$ are the eigenvalues of $\rho_A$, with $\lambda_0 \geq ... \geq \lambda_{\alpha-1}$ and $\sum_{i=0}^{\alpha-1}\lambda_i = 1$. When $d \rightarrow 1$, $E_d$ is the von Neumann entropy $-\sum_{i=0}^{\alpha-1}\lambda_i\ln\left(\lambda_i\right)$ \cite{nielsen2011r}, when $d = 2$, $E_d$ is the R\'{e}nyi entropy or the Collision entropy, and when $d \rightarrow \infty$, it is the Minimal entropy $-\ln\left(\lambda_0\right)$.

The R\'{e}nyi entropy $E_d$ for $d\rightarrow1$, $d=2$ and $d\rightarrow\infty$ can be calculated analytically by integrating over the MPD using the Eq. (\ref{eq_A4}) and expressed in terms of $\alpha$ and $\lambda_0$

\begin{widetext}

\begin{equation}
\int^b_0 \ln\left(x\right)\sqrt{x\left(b-x\right)}dx=\frac{\pi}{16}\left(2b^2\ln\left(b\right)-b^2\left(4\ln\left(2\right)-1\right)\right),
\label{eq_A4}
\end{equation}

\begin{equation}
\begin{aligned}
\lim_{d\rightarrow1}E_d&=-\frac{\alpha}{2\pi\sigma_\gamma^2} \int_0^{4\sigma_\gamma^2}\ln\left(x\right)\sqrt{\left(4\sigma_\gamma^2 -x\right)x}dx-\lambda_0\ln\left(\lambda_0\right)\\
&=\left(1-\lambda_0\right)\ln\left(\frac{\alpha}{4\left(1-\lambda_0\right)}\right)-\frac{\lambda_0}{2}\left(4\ln\left(2\right)-1\right)-\lambda_0\ln\left(\lambda_0\right)+2\ln\left(2\right)-\frac{1}{2},
\end{aligned}
\label{eq_renyi_1}
\end{equation}
\end{widetext}

\begin{equation}
\begin{aligned}
E_2&=-\ln\left(\frac{\alpha}{2\pi\sigma_\gamma^2}\int_0^{4\sigma_\gamma^2}x\sqrt{\left(4\sigma_\gamma^2 -x\right)x}dx+\lambda_0^2\right)=-\ln\left(\frac{\lambda_0^2(\alpha+2)-4\lambda_0+2}{\alpha}\right),
\end{aligned}
\end{equation}

and

\begin{equation}
\lim_{d\rightarrow\infty}E_d=-\ln\left(\lambda_0\right).
\end{equation}

\begin{figure}[h]
\centering
\includegraphics[width=0.5\linewidth]{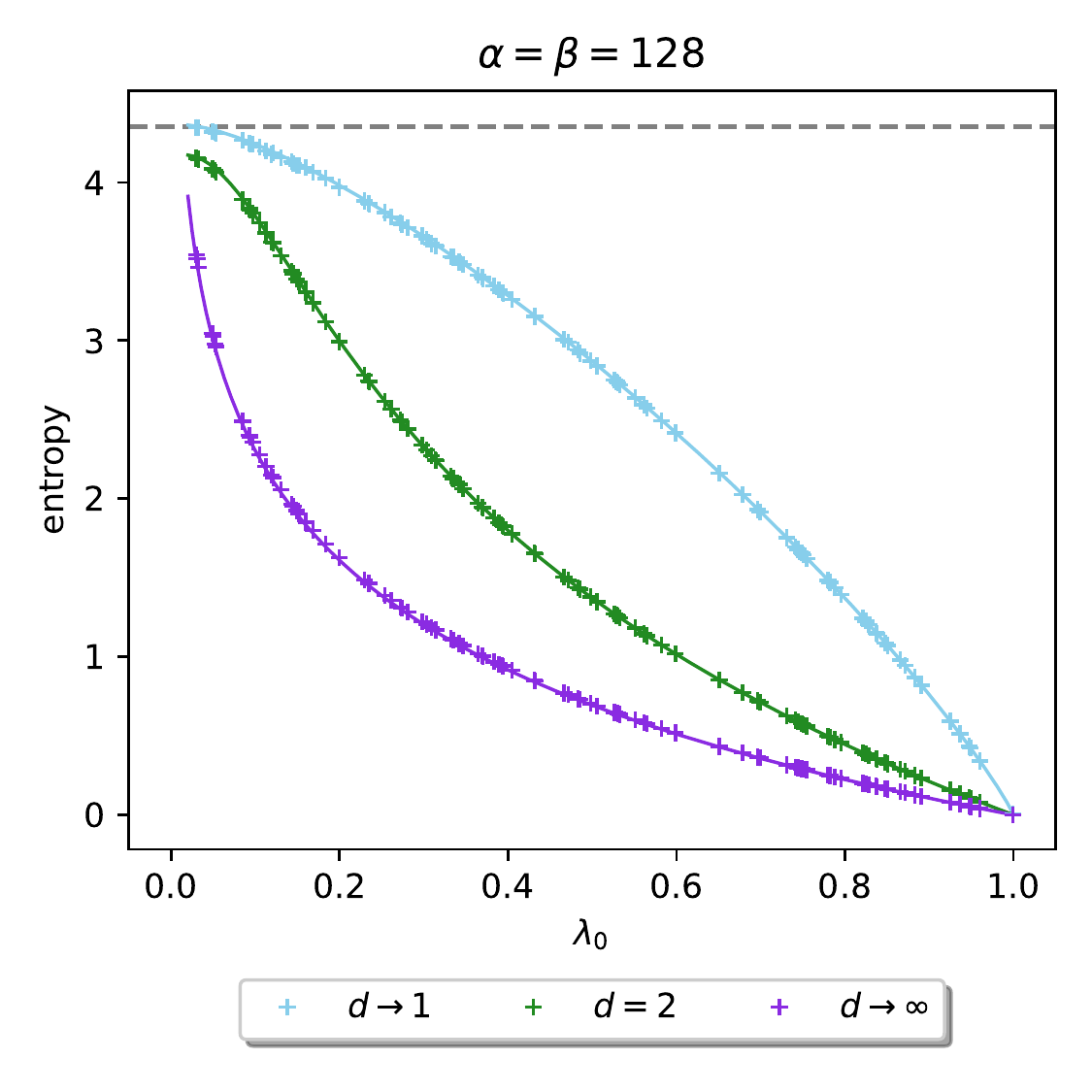}
\caption{The numerical simulation (in $+$) versus the analytical solution (in line). The R\'{e}nyi entropy for $d\rightarrow0$ (here we take $d=1.001$), $d=2$ and $d\rightarrow\infty$ (here we take $d=100$) for $100$ normalized random reduced density matrices (same matrices for each $d$) and the analytic results. The gray dotted line is $\ln\left(\alpha\right)-\frac{1}{2}$.}
\label{Renyi}
\end{figure}

These three equations are verified with numerical results in FIG. \ref{Renyi}. We can recover the Eq. (\ref{eq_renyi_1}) from the Eq. (\ref{eq_E_alpha_beta_lambda_3}) by setting $\beta=\alpha$. R\'{e}nyi entropy with other degrees can be obtained using integrals in the TABLE. \ref{theory}. Appendix \ref{app_adiabatic} and \ref{app_prime} provide physical examples of this analysis.

\begin{table}
\begin{tabular}{ |c|c|c|c|c|c| } 
\hline
$d$ & 2 & 3 & 4 & 5 & 6\\ 
\hline
$\int_0^{t}x^{d-1}\sqrt{\left(t-x\right)x}dx$ & $\frac{1}{16}\pi t^3$ & $\frac{5}{128}\pi t^4$ & $\frac{7}{256}\pi t^5$ 
& $\frac{21}{1024}\pi t^6$ & $\frac{33}{2048}\pi t^7$\\
\hline
\end{tabular}
\caption{Integration table of $\int_0^{t}x^{d-1}\sqrt{\left(t-x\right)x}dx$.}
\label{theory}
\end{table}

\section{Conclusion}

We postulate the behavior of decentralized Wishart matrix eigenvalues from the Marchenko Pastur distribution and the spectral norm of matrices. We derive a precise analytical link between the dominant eigenvalue and various entropy of the reduced density matrix. For example, the von Neumann entropy obtained by tracing various subsystems of a random pure state and different degrees of R\'{e}nyi entropy of a bi-partition system is included in the set of entropy. We also investigate the entanglement gap and derive a decreasing relation between the entropy and the entanglement gap. Although the exact result is obtained with the study of Wishart matrices, its application can be extended to more general cases. Examples of physical systems of quantum computing are mentioned in the appendices.

\begin{acknowledgments}
The author would like to thank Prof. J. I. Latorre and Prof. G. Sierra for their helpful information and discussion. Also, the author is grateful for a mathematical hint from one anonymous reviewer.
\end{acknowledgments}

\appendix

\section{QFT of random states}\label{app_QFT}

The analytical result in Eq. (\ref{eq_E_alpha_beta_lambda_3}) remains valid for the Quantum Fourier Transform (QFT) \cite{nielsen2002quantum,weinstein2001implementation}. Considering a random state $\ket{\psi}=\sum_{i=1}^{\alpha\beta}c_i\ket{i}$ with $\sum_{i=1}^{\alpha\beta}|c_i|^2=1$. Assuming that the dominant eigenvalues of its reduced density matrix $\rho_A$ is still $\lambda_0$, then the normalized state after performing the QFT on $\ket{\psi}$ can be written as

\begin{equation}
QFT\ket{\psi}=Q_0\ket{0}+\sum_{i=1}^{N-1}Q_i\ket{i},
\end{equation}

where $|Q_0|^2\sim\lambda_0$ and $Q_i$ are random variables with mean $0$ and variance $\sigma_{QFT}^2=\left(1-\lambda_0\right)/\alpha\beta$, which according to the Central limit theorem \cite{rosenblatt1956central}, can be approximated by $\mathcal{N}_{\mathbb{C}}\left(0,\sigma_{QFT}\right)$. The reduced density matrix of $QFT\ket{\psi}$ can be represented under the form

\begin{equation}
\rho_{QFT}=
\begin{pmatrix}
q_0 & q_2 & ... & q_2 & q_2\\
q_2 & q_1 & ... & q_3 & q_3\\
\vdots & \vdots & & \vdots & \vdots\\
q_2 & q_3 & ... & q_1 & q_3\\
q_2 & q_3 & ... & q_3 & q_1
\end{pmatrix},
\label{eq_Q}
\end{equation}

where

1) $q_0\sim\lambda_0$,

2) $q_1\sim\mathcal{N}\left(\left(1-\lambda_0\right)/\alpha,\sigma_1\right)$ with $\sigma_1\sim\mathcal{O}\left(\alpha\sigma_{QFT}^2\right)$,

3) $q_2\sim\mathcal{N}_\mathbb{C}\left(0,\sigma_2\right)$ with $\sigma_2\sim\sqrt{\lambda_0}\sigma_{QFT}$,

4) and $q_3\sim\mathcal{N}_\mathbb{C}\left(0,\sigma_3\right)$ with $\sigma_3\sim\mathcal{O}\left(\alpha\sigma_{QFT}^2\right)$.

\begin{figure}
\centering
\includegraphics[width=0.5\linewidth]{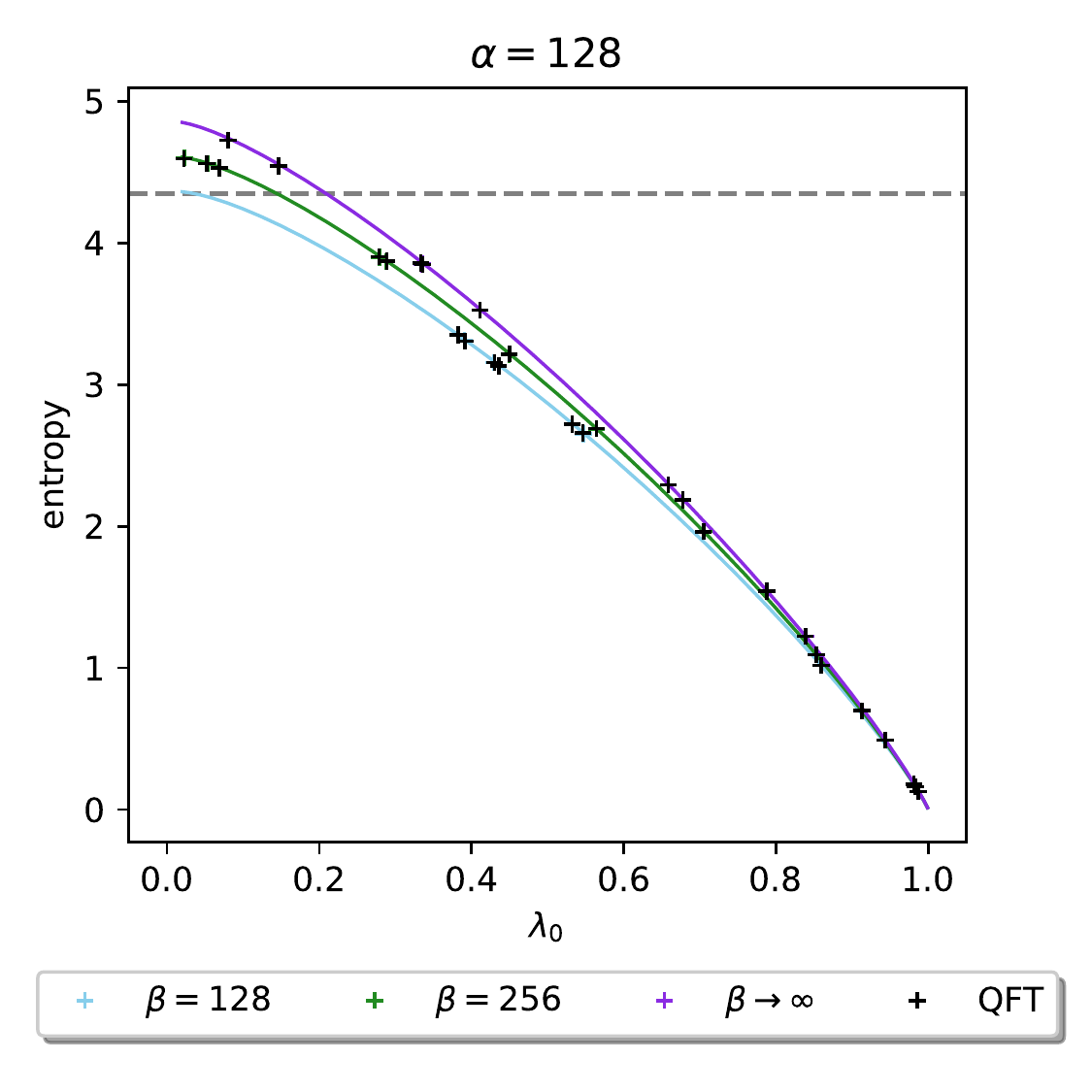}
\caption{The von Neumann entropy before and after applying the QFT of different $\beta>\alpha$ for the same $\alpha=128$, $10$ random states for each $\beta$. Notice that data points before and after applying the QFT overlap completely. Here for the numerical result of $\beta\rightarrow\infty$ we use $\beta=8192$. The gray dotted line is $\ln\left(\alpha\right)-\frac{1}{2}$.}
\label{QFT}
\end{figure}

The symbol $\mathcal{O}$ is used to give an order of approximation since the elements of $QFT\ket{\psi}$ are correlated and can not be treated as i.d.d random variables. When $\alpha\gg 0$, for a given $\lambda_0$, we have $q_0\gg q_1,q_2,q_3$, which indicate that $\lambda_0^{QFT}$ the dominant eigenvalues of $\rho_{QFT}$ is very likely to be near $\lambda_0$. We can observe from FIG. \ref{QFT} that $QFT\ket{\psi}$ and $\ket{\psi}$ share the exact dominant eigenvalue and entropy. This result holds true for random quantum states, which is different from the usual case, where the QFT creates entanglement \cite{mastriani2021quantum}. 

\section{Grover algorithm to solve a toy Hash function}\label{app_grover}

\begin{figure}
\centering
\includegraphics[width=0.5\linewidth]{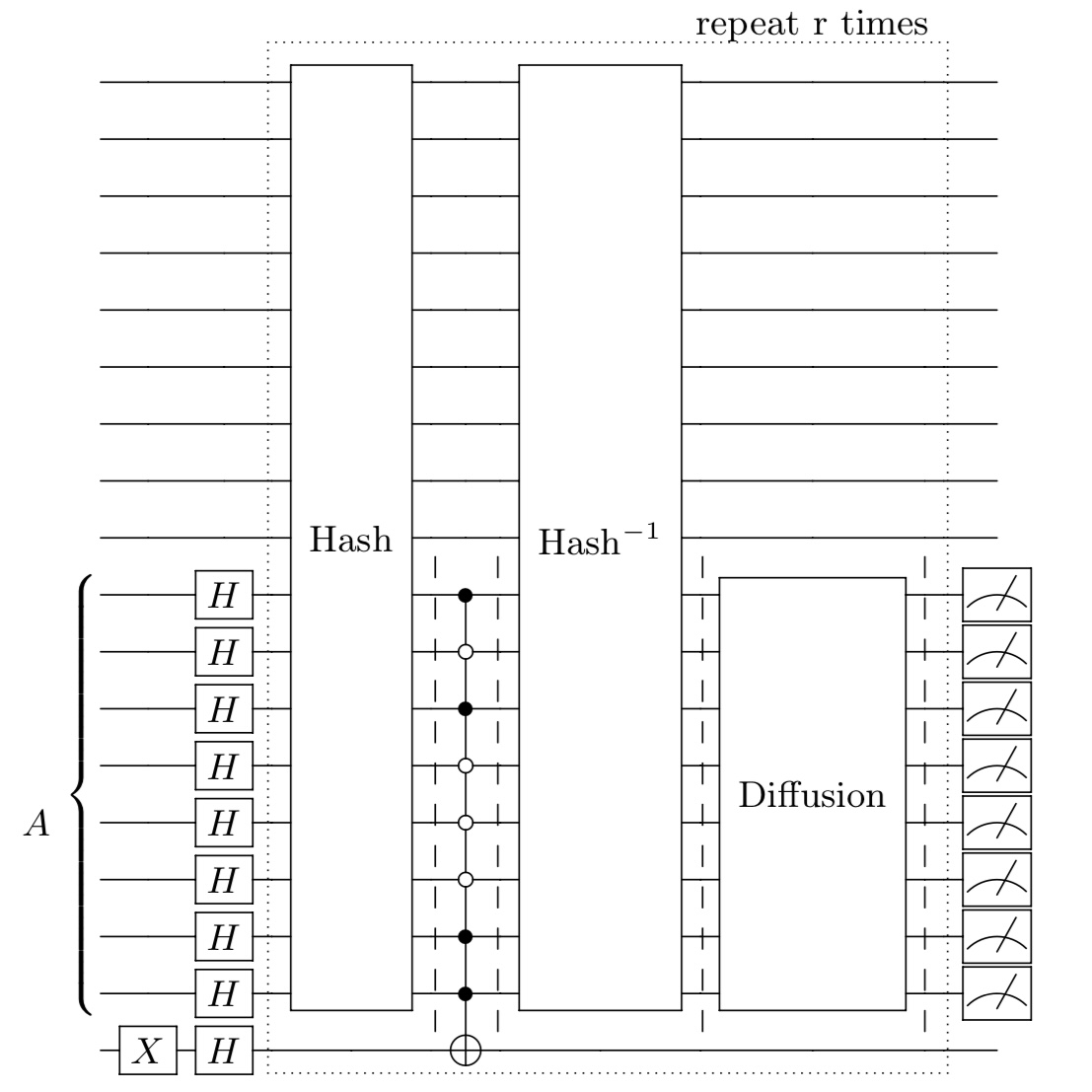}
\caption{The quantum circuit of a toy Grover's algorithm to find the pre-image of a Hash function with a given cipher-text. The oracle consists of a unitary that encodes the Hash function, the inverse of this unitary, with a multi-Toffoli gate in the middle that encodes the cipher-text, which is $10100011$ in our example. The $8$-qubit subsystem that we study is indicated as $A$. We sample the reduced density matrix in each iteration after the Hash unitary, the multi-Toffoli gate, the inverse Hash unitary, and the diffusion operator as indicated with dashed lines. The cipher-text that we choose has $2$ pre-images, and the number of Grover iterations needed is $r=8$.}
\label{Grover_circuit}
\end{figure}

\begin{figure}
\centering
\includegraphics[width=0.5\linewidth]{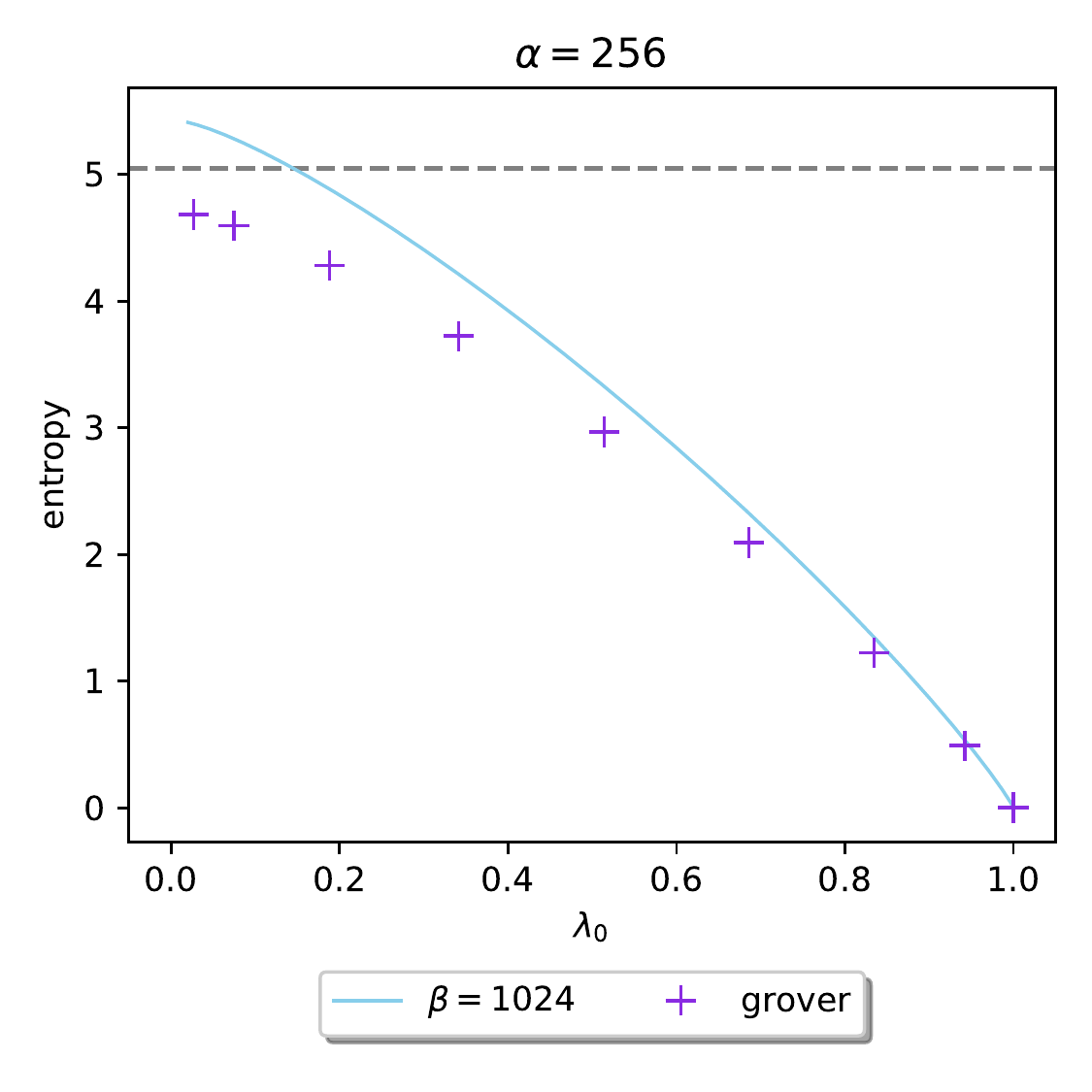}
\caption{The entanglement entropy of the quantum search circuit to solve the pre-image of a Hash function versus the analytical result in the Eq. (\ref{eq_E_alpha_beta_lambda_3}). The gray dotted line is $\ln\left(\alpha\right)-\frac{1}{2}$. Notice that the logarithm is in base $2$ in the original paper, and we use natural logarithm in this article.}
\label{Grover}
\end{figure}

In Ref. \cite{ramos2021quantum}, the authors have implemented Grover’s algorithm in a quantum simulator Qibo \cite{efthymiou2021qibo} to perform a quantum search for pre-images of a toy Sponge Hash function \cite{bernstein2008chacha}. Their circuit consists of $18$ qubits, where $8$ of them are initialized as uniform superposition of all states, refer to as the search space of Grover's algorithm, details of are provided in their paper and their GitHub page \cite{Github_hash}. In this article, we are only interested in tracking the entanglement evolution of an actual quantum algorithm and use their implementation as an example. A simplified circuit is shown in FIG. \ref{Grover_circuit}. We study the reduce density matrices $\rho_A$, where $A$ is the search space and the complementary space is the rest of the qubits. Results are shown in FIG. \ref{Grover}. The relation between the dominant eigenvalue and the von Neumann entropy of these reduced density matrices shares a similar pattern as predicted by our research on Wishart matrices. As randomness represented maximum unpredictability, it is natural that the entanglement of a physical system is inferior to the analytical solution.

\section{Adiabatic quantum computation to solve the Exact Cover problem}\label{app_adiabatic}

In Ref. \cite{orus2004universality}, the authors have studied the entanglement properties of the adiabatic quantum algorithm for solving the NP-complete Exact Cover (EC) problem, which is a particular case of $3$-SAT problem. In a nutshell, the quantum system is initialized with an "easy" Hamiltonian $H_0$, which is a magnetic field in the $x$ direction, where the ground state is an equal superposition of all possible computational states. The system is then adiabatically evolved to a "complicated" Hamiltonian $H_p$ that encodes an instance of EC problem, where the ground state is the solution. During the adiabatic evolution, the system can be described using a Hamiltonian $H_s$, an interpolation between $H_0$ and $H_p$,

\begin{equation}
H_s:=\left(1-s\right)H_0+sH_p\hspace{0.5cm}\text{with}\hspace{0.5cm}s\in[0,1].
\end{equation}

\begin{figure}
\centering
\includegraphics[width=0.5\linewidth]{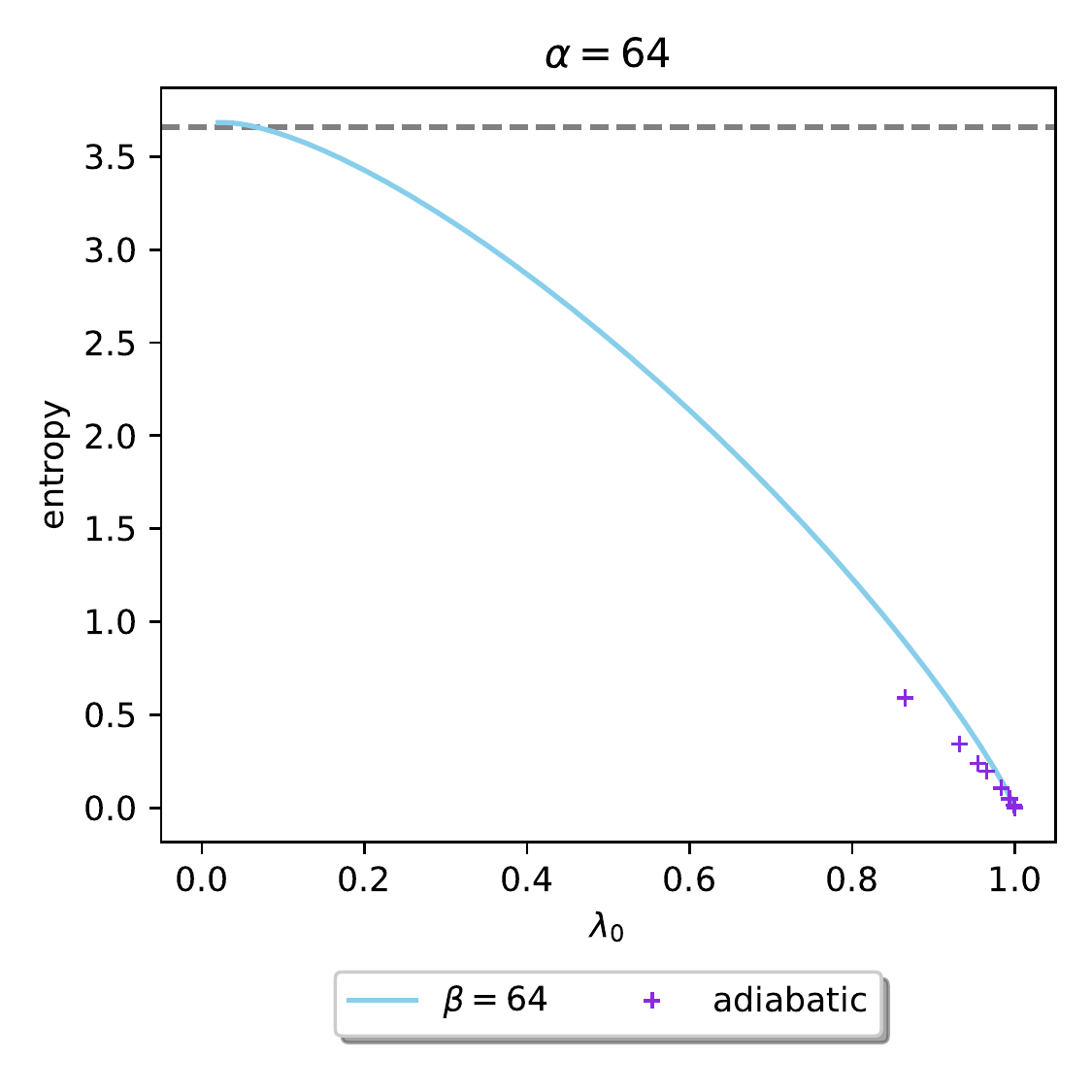}
\caption{The entanglement entropy of the ground state of $H_s$, which solves one instance of EC problem with $n=12$ versus the analytical result in the Eq. (\ref{eq_renyi_1}). The gray dotted line is $\ln\left(\alpha\right)-\frac{1}{2}$. Notice that the logarithm is in base $2$ in the original paper, and we use natural logarithm in this article.}
\label{EC}
\end{figure}

For a given $12$-qubit instance of the EC problem, we track the von Neumann entropy of the natural bi-partition (the first half of the qubits is traced over the second half of the qubits) during one adiabatic evolution. The result is shown in FIG. \ref{EC}. The entanglement of the ground state of $H_s$ is upper-bounded by and shares a similar feature with the analytical result that we deduced with RMT. It is known that quantum states whose von Neumann entropy only scales logarithmically with the number of qubits can be described in terms of Matrix Product States \cite{vidal2003efficient}. This means the amount of entropy present in the system along the quantum algorithm should be large; otherwise, an efficient algorithm would exist to solve the same problem using Tensor Networks. The dominant eigenvalue is easier to obtain than the rest with certain algorithms \cite{lanczos1950iteration,chen2010effect}. Therefore, our result can be useful for estimating the advantage of a given quantum algorithm.

\section{Prime state}\label{app_prime}

In Ref. \cite{latorre2013quantum,latorre2014there,garcia2020prime}, the authors have studied the entanglement of the Prime state. Prime state is defined as an equally weighted superposition of prime numbers in $n$ qubits computational basis

\begin{equation}
\ket{\mathbb{P}_n}:=\frac{1}{\sqrt{\pi\left(2^n\right)}}\sum_{p\in\text{prime}\leq 2^n}\ket{p},
\end{equation}

where each prime number $p=p_0 2^0 + p_1 2^1 + ... + p_{n-1} 2^{n-1}$ is implemented as $\ket{p}=\ket{p_{n-1},...,p_{0}}$, and $\pi\left(2^n\right)$ is the amount of prime numbers less than $2^n$. 

The bi-partition entanglement entropy of the Prime state is investigated in \cite{latorre2014there}. Here we explore further if it fits our analysis.

Let $\{\rho^{\mathbb{P}_n}_{A_k}\}_{k \in \{1,...,\binom{n}{n/2}/2\}}$ be the reduced density matrix set of all possible bi-partitions of $\ket{\mathbb{P}_n}$, where $\binom{n}{n/2}/2$ is the binomial coefficient of choosing $n/2$ qubits out of $n$. It is divided by $2$ since the reduced density matrix is identical when tracing over one subsystem and its complementary system. 

\begin{figure}
\centering
\includegraphics[width=0.5\linewidth]{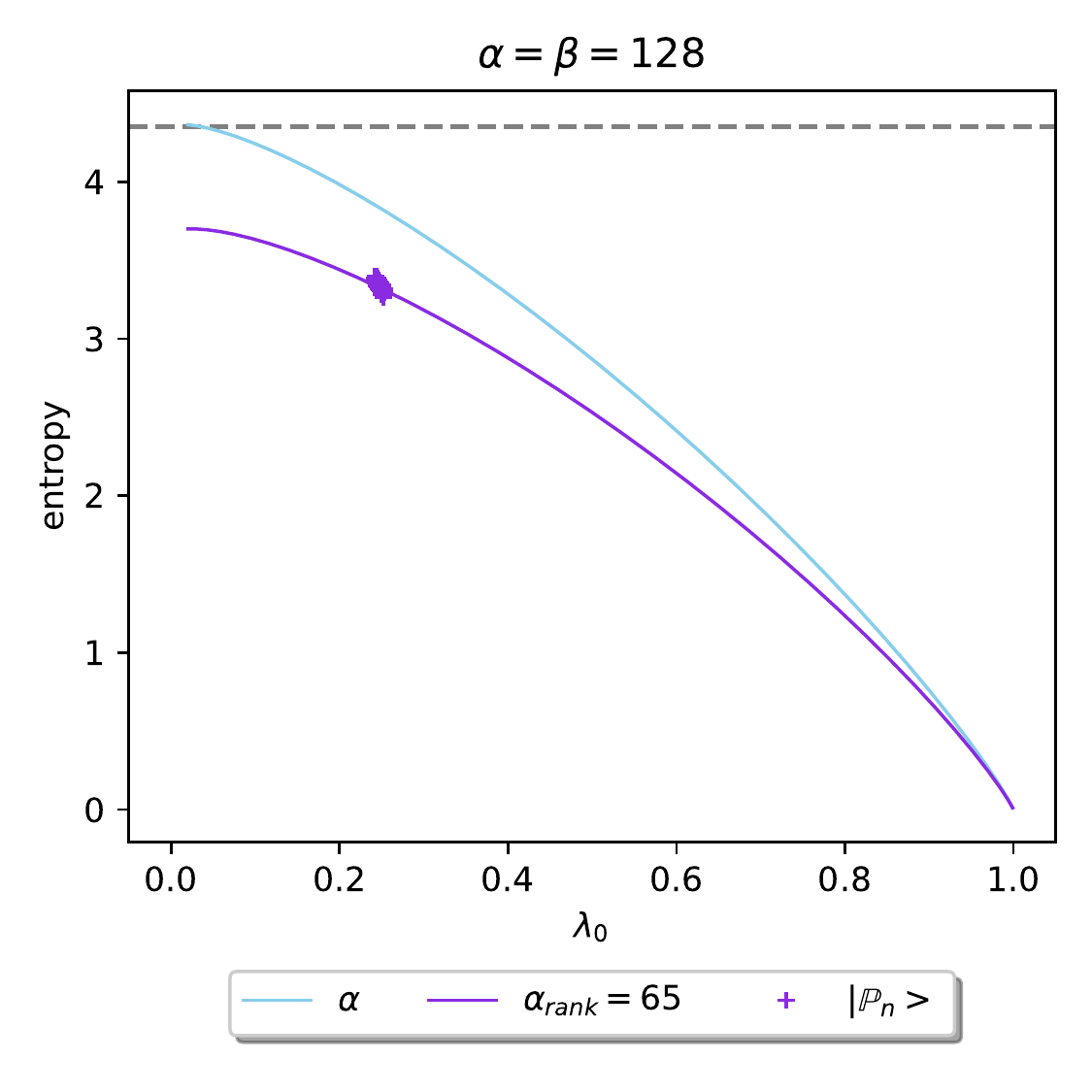}
\caption{The entanglement entropy of all possible bi-partitions of Prime state with $n=14$ versus the analytical result in Eq. (\ref{eq_renyi_1}). The gray dotted line is $\ln\left(\alpha\right)-\frac{1}{2}$. Notice that the logarithm is in base $2$ in the original paper, and we use natural logarithm in this article.}
\label{Prime}
\end{figure}

In this article, only full-rank random matrices have been considered. However, since all prime numbers are odd except the number $2$, every $\rho^{\mathbb{P}_n}_{A_k}$ has the same rank $\alpha_{rank}=2^{\frac{n}{2}-1}+1$, which also denotes as the Schmidt rank or the Schmidt number \cite{vidal2003efficient,sperling2011schmidt,nielsen2002quantum}. The result needs to be adapted accordingly. The von Neumann entropy and the dominant eigenvalue of $\rho^{\mathbb{P}_n}_{A_k}$ is located near the line defined by the Eq. (\ref{eq_renyi_1}) replacing $\alpha$ with $\alpha_{rank}$ as shown in the FIG. \ref{Prime}.

\end{document}